\pdfoutput=1
\documentclass[%
 reprint,
superscriptaddress,
 amsmath,amssymb,
 aps,
floatfix,
]{revtex4-2}

\usepackage{graphicx}
\usepackage{dcolumn}
\usepackage[T1]{fontenc} 
\usepackage{siunitx}
\usepackage{verbatim}
\DeclareSIUnit{\belmilliwatt}{Bm}
\DeclareSIUnit{\e}{e}
\DeclareSIUnit{\dBm}{\deci\belmilliwatt}
\usepackage{amssymb}
\usepackage{xcolor}
\usepackage{braket}
\usepackage{bm}
\usepackage{booktabs}
\usepackage{hyperref}

\begin{document}

\preprint{APS/123-QED}

\title{Cryogenic hyperabrupt strontium titanate varactors for  sensitive reflectometry of quantum dots}

\author{Rafael~S.~Eggli}
\email{e-mail: rafael.eggli@unibas.ch, dominik.zumbuhl@unibas.ch}
\affiliation{Department of Physics, University of Basel, Klingelbergstrasse 82, \\ CH-4056 Basel, Switzerland}

\author{Simon~Svab}
\thanks{This author contributed equally to this work as the first author}
\affiliation{Department of Physics, University of Basel, Klingelbergstrasse 82, \\ CH-4056 Basel, Switzerland}


\author{Taras~Patlatiuk}
\affiliation{Department of Physics, University of Basel, Klingelbergstrasse 82, \\ CH-4056 Basel, Switzerland}

\author{Dominique~A.~Tr\"ussel}
\affiliation{Department of Physics, University of Basel, Klingelbergstrasse 82, \\ CH-4056 Basel, Switzerland}

\author{Miguel~J.~Carballido}
\affiliation{Department of Physics, University of Basel, Klingelbergstrasse 82, \\ CH-4056 Basel, Switzerland}

\author{Pierre~Chevalier~Kwon}
\affiliation{Department of Physics, University of Basel, Klingelbergstrasse 82, \\ CH-4056 Basel, Switzerland}

\author{Simon~Geyer}
\affiliation{Department of Physics, University of Basel, Klingelbergstrasse 82, \\ CH-4056 Basel, Switzerland}

\author{Ang~Li}
\thanks{Current address: Institute of Microstructure and Properties of Advanced Materials, Beijing University of Technology, Beijing, 100124, China.}
\affiliation{Department of Applied Physics, TU Eindhoven, Den Dolech 2, 5612 AZ Eindhoven, The Netherlands}

\author{Erik~P.~A.~M.~Bakkers}
\affiliation{Department of Applied Physics, TU Eindhoven, Den Dolech 2, 5612 AZ Eindhoven, The Netherlands}

\author{Andreas~V.~Kuhlmann}
\affiliation{Department of Physics, University of Basel, Klingelbergstrasse 82, \\ CH-4056 Basel, Switzerland}

\author{Dominik~M.~Zumb\"uhl}
\email{e-mail: rafael.eggli@unibas.ch, dominik.zumbuhl@unibas.ch}
\affiliation{Department of Physics, University of Basel, Klingelbergstrasse 82, \\ CH-4056 Basel, Switzerland}

\begin{abstract}
Radio frequency reflectometry techniques enable high bandwidth readout of semiconductor quantum dots. Careful impedance matching of the resonant circuit is required to achieve high sensitivity, which however proves challenging at cryogenic temperatures. Gallium arsenide-based voltage-tunable capacitors, so-called varactor diodes, can be used for in-situ tuning of the circuit impedance but deteriorate and fail at temperatures below \SI{10}{\kelvin} and in magnetic fields. Here, we investigate a varactor based on strontium titanate with hyperabrupt capacitance-voltage characteristic, that is, a capacitance tunability similar to the best gallium arsenide-based devices. The varactor design introduced here is compact, scalable and easy to wirebond with an accessible capacitance range from \SIrange{45}{3.2}{\pico\farad}. We tune a resonant inductor-capacitor circuit to perfect impedance matching and observe robust, temperature and field independent matching down to \SI{11}{\milli\K} and up to \SI{2}{\tesla} in-plane field. Finally, we perform gate-dispersive charge sensing on a germanium/silicon core/shell nanowire hole double quantum dot, paving the way towards gate-based single-shot spin readout. Our results bring small, magnetic field-resilient, highly tunable varactors to \SI{}{\milli\K} temperatures, expanding the toolbox of cryo-radio frequency applications.

\end{abstract}

\maketitle


\section{Introduction}
Recent years have seen rapid progress in the realization of spin qubits in semiconductors \cite{Loss1998,Hendrickx2021,Takeda2022,Philips2022}. Hole spins in germanium (Ge) and silicon (Si) nanostructures are of particular interest thanks to the availability of all-electrical spin control \cite{Maurand2016,Kloeffel2011,Kloeffel2018,Froning2021,Geyer2021,Hendrickx2021}, short gate times \cite{Froning2021a,Geyer2022,Wang2022}, high-temperature qubit operation \cite{Camenzind2022,Petit2020,Yang2020}, intrinsically low concentration of spinful nuclei and reduced hyperfine interaction due to the p-type hole wavefunction \cite{Fang2022} and the presence of noise sweet spots \cite{Bosco2021,Bosco2021a,Piot2022}. 

Fast readout of charge and spin states in semiconductor qubit devices is typically achieved using high-frequency reflectometry methods such as radio frequency (RF)-single-electron transistors \cite{Reilly2007,Angus2008,Barthel2009,Takeda2016}, single lead sensor dots \cite{House2016,Oakes2023} or gate-dispersive charge sensing \cite{Colless2013,GonzalezZalba2015,West2019,Vigneau2022,Urdampilleta2019}. The latter constitutes the smallest footprint charge sensing setup: The same gate that electrostatically defines a quantum dot is used to sense nearby dots. This makes gate-reflectometry the obvious choice to use with nanowire-qubits since external charge sensors are difficult to implement \cite{Hu2007,Hu2011,Jung2012,Higginbotham2014,Ungerer2022}. 

High device impedances, typically on the order of the resistance quantum  $Z\geq\SI{25.8}{\kilo\ohm}$ up to $Z\sim\SI{}{\giga\ohm}$ for gate-sensors make it necessary to downconvert towards  $Z_0=\SI{50}{\ohm}$ for maximizing signal and bandwidth. Downconversion is implemented using inductor-capacitor (LC) resonators, so-called tank circuits, consisting of surface-mount inductors and parasitic circuit or device capacitances \cite{Vigneau2022}. Careful circuit design ensures optimal signal-to-noise ratios (SNRs) and high sensitivities \cite{GonzalezZalba2015,Ahmed2018}. The RF characteristics of the circuits are frequently affected by cryogenic temperatures and magnetic fields required for qubit operations. Thus, in-situ tuning of the impedance matching and thereby the coupling of the RF source circuitry and the LC resonator is needed. This can be achieved using varactors \cite{Mueller2010,Ares2016,House2016,Ibberson2019,Fang2022}. 

Commercial varactors with high tuning range and steep capacitance-voltage (CV)-curves, classified as hyperabrupt varactors, are typically doped gallium arsenide (GaAs) diodes. The width of the diode's p-n junction depletion region can be tuned by applying a reverse bias voltage, effectively changing the capacitor thickness and thus the diode's series capacitance. Doped varactors are suitable for quantum applications only to a limited extent, requiring multiple diodes \cite{Ares2016,Ibberson2019}. This is due to freezing-out effects \cite{Ibberson2019} and strong dependencies on applied magnetic fields. In particular, perfect impedance matching using GaAs varactors was so far only demonstrated down to $T\sim\SI{200}{\milli\kelvin}$, using two varactors \cite{Ibberson2019}. 

To resolve these issues, quantum paraelectric materials such as strontium titanate (STO) are under consideration as alternative cryogenic varactors \cite{Johnson1962,Apostolidis2020}. Below \SI{10}{\kelvin}, the dielectric constant of STO ranges from $\epsilon_r=10'000-24'000$ depending on the crystal orientation and was shown to be tunable with electric fields, reaching  more than a factor of 10 lower dielectric constants    \cite{Saifi1970,Sakudo1971,Neville1972,Hemberger1995}. This strong electric field tunability makes STO a prime candidate for the design of tunable cryogenic RF components, such as varactors or resonators \cite{Davidovikj2017,Apostolidis2020, Johnson1962}. The relatively small target capacitance range $C_{var}<\SI{50}{\pico\farad}$ \cite{Ares2016} required for impedance matching makes STO varactor engineering challenging. Given the high $\epsilon_r$, a plate capacitor formed by electrodes on opposite faces of STO crystals encounters conflicting constraints when required to reach such low $C_{var}$: i) either the capacitor plate size is too small for wirebonding or ii) the plate-to-plate distance is too large to form uniformly high fields, reducing the tunability \cite{Apostolidis2020}. Furthermore, contacting the electrodes on opposing faces of the crystal is inconvenient and limits scalability.

Here, we present a surface-patterned ring varactor design on STO which achieves hyperabrupt CV characteristics and a tuning range exceeding that of typical GaAs varactors \cite{Ares2016,House2016,Ibberson2019} while operating down to \SI{}{\milli\K} temperatures and independent of magnetic fields. We fabricate varactors of various dimensions and characterize them first at \SI{4}{\kelvin}, optimizing for tunability and capacitance range. 

Next, we integrate such a varactor in a tank circuit for gate-dispersive readout. The selected design has a \SI{300}{\micro\meter} footprint, smaller than STO plate-capacitor layouts and packaged commercial varactor diodes. Critical coupling is reached over a broad range of temperatures from $T\approx\SI{4}{\K}$ to \SI{11}{\milli\K} to and in magnetic fields up to \SI{2}{\tesla} without retuning of the varactor. We use the matched circuit to dispersively measure charge transitions in a hole double quantum dot (DQD) formed in a Ge/Si core/shell nanowire (NW) \cite{ConesaBoj2017,Froning2018,Hao2010,Xiang2006}. Similar NWs were used to demonstrate all-electrical, ultrafast spin qubit operation with a high degree of tunability \cite{Kloeffel2011,Kloeffel2018,Froning2021,Froning2021a}. To directly probe the significance of impedance matching for gate-dispersive charge sensing, we measure signal-to-noise ratios (SNRs) for variable matching conditions and at different charge transitions. Our implementation of gate-dispersive charge sensing paves the way for high-fidelity gate-dispersive single-shot spin readout with gated quantum dots (QDs). The realization of a highly tunable capacitor at cryogenic temperatures enables more sophisticated circuit architectures in various applications beyond charge sensing in semiconductors.

\section{Varactor Characterization at 4~K}

\subsection{Geometry}
\label{sect:geometry}

\begin{figure}[hbtp]
    \includegraphics[width=0.47\textwidth]{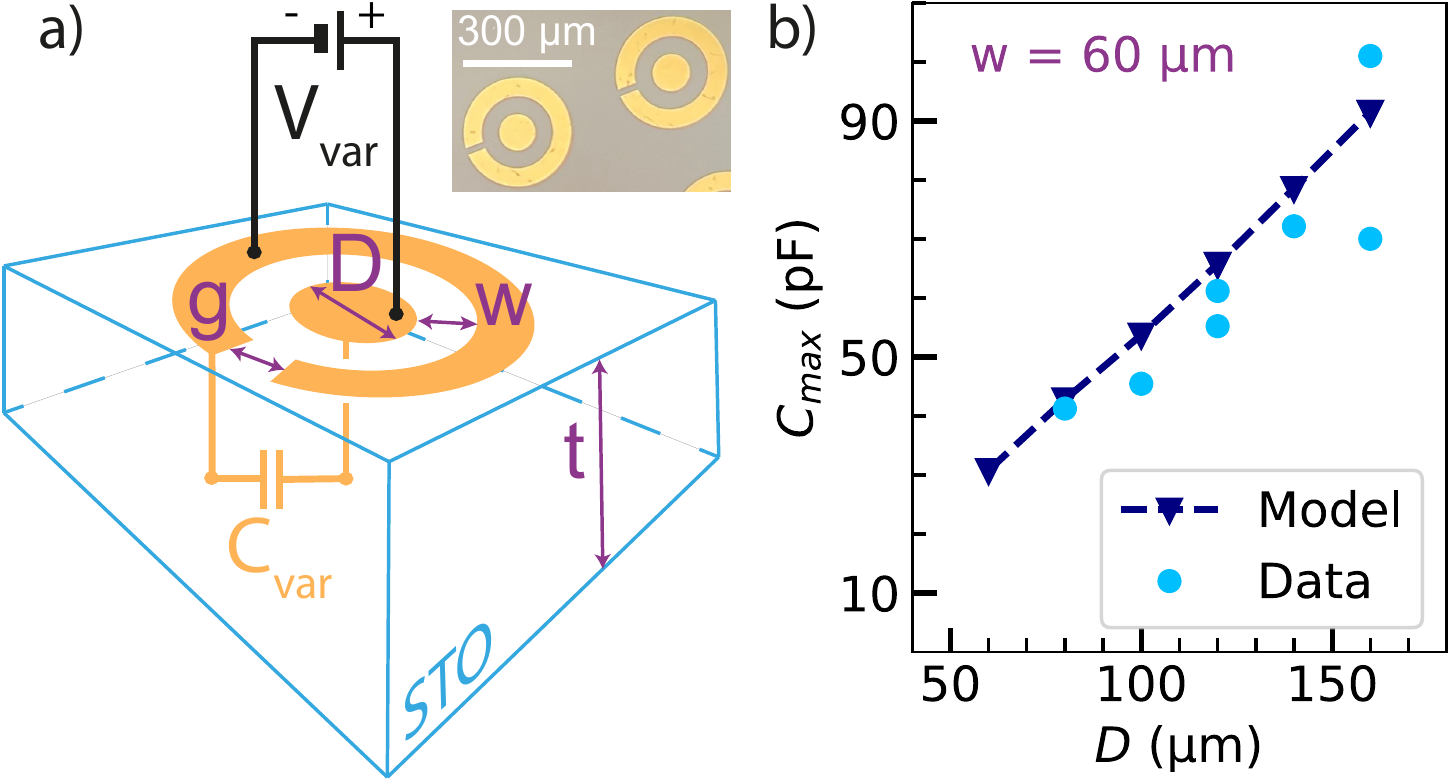}
    \caption{\textbf{Geometry of STO ring varactors} a) Schematic diagram of the STO varactor design. Two gold electrodes, a circle with diameter $D$ and a concentric ring with a spacing to the circle $w$ and gap $g=\SI{20}{\micro\meter}$ form the capacitor. Application of a DC voltage $V_{var}$ induces an electric field within the STO crystal which is highly localized due to the compact geometry. Inset: Optical micrograph of finished varactors. b) Zero-field capacitance $C_{max}$ as a function of $D$ for $w=\SI{60}{\micro\meter}$, simulated in COMSOL (triangles) and measured (circles).}
    \label{fig:Geometry}
\end{figure}

The varactor ring design is presented in Fig. \ref{fig:Geometry} a). It features two electrodes, a central circle with diameter $D$ and a concentric ring, \SI{50}{\micro\meter} wide and separated from the circle by a distance $w$. A gap $g=\SI{20}{\micro\meter}$ wide in the outer ring facilitates lift-off. The electrodes were patterned on the polished side of a commercially available $(001)$ STO chiplet with thickness $t=\SI{0.5}{\milli\meter}$ (Appendix \ref{Fab:Var}). This geometry provides several key features for optimal varactor performance: electric fields in the crystal are localized and large close to the electrodes. This allows for high tunability of $\epsilon_r$ while keeping the peak capacitance $C_{max}$ in the desired range. The small varactor footprint of $\sim\SI{300}{\micro\meter}$ and single-sided metallization allows tight integration with other circuit components and provides scalability. In particular, we are able to fit 9 independent ring varactors on a chiplet the size of commercial surface-mount GaAs varactors used in comparable experiments (\textit{MACOM} MA46-series) \cite{Ares2016,House2016,Ibberson2019}.

All measurements presented in this section were performed at \SI{4}{\kelvin} using standard lock-in techniques at \SI{1}{\kilo\hertz}. No frequency dependence of tunability and $\epsilon_r$ from \SI{1}{\kilo\hertz} to \SI{50}{\mega\hertz} was reported previously, predicting constant $\epsilon_r$ up to \SI{100}{\giga\hertz} \cite{Neville1972}. We thus expect our low-frequency results to apply at typical reflectometry frequencies as well. We estimate the zero-field capacitance $C_{max}$ of our design using \textit{COMSOL} to find optimal device dimensions. For simplicity, we assume an isotropic dielectric constant $\epsilon_r=24'000$, approximately the experimental value for field lines parallel to the $(001)$ direction at \SI{4}{\kelvin} \cite{Neville1972}. In our simulation, we primarily focus on $D$ and $w$, finding that neither $g$ nor the width of the outer ring have a strong influence on $C_{max}$. We vary $D$ from \SIrange{60}{160}{\micro\meter}  and $w$ from \SIrange{10}{80}{\micro\meter} and fabricate an array of 36 different combinations, excluding $w>\SI{60}{\micro\meter}$ due to decreased tunability. Not all combinations were measured because too small $w$ and $D$ frequently resulted in shorts between the two electrodes from wirebonding (see Appendix \ref{Supp_sect_geometry} for details).

In Fig. \ref{fig:Geometry} b), we compare the zero-field capacitance $C_{max}$ estimated from finite-element simulations and measurements for $w=\SI{60}{\micro\meter}$ and variable $D$. We find good agreement, noting that some small capacitance over-estimations by the model could be explained by the anisotropy of $\epsilon_r$, which was neglected in the simulation. Furthermore, the partial loss of tunability caused by the application of negative varactor voltages $V_{var}$ described in \ref{sect:tunability} and Fig. \ref{fig:CV-Characteristic} c) may also explain some of the measured $C_{max}$ lying below the model, as some of these values were taken from voltage sweeps starting at a varactor voltage $V_{var}=\SI{-2}{\volt}$.

\subsection{Voltage-Tunability}
\label{sect:tunability}

The capacitance-tunability of STO varactors relies on an incipient ferroelectric phase transition at low temperatures which is never reached due to quantum fluctuations \cite{Ang2004,Davidovikj2017, Apostolidis2020,Johnson1962}. Below about 10~K, the low temperature limit is reached and $\epsilon$ has no further temperature dependence, thus rendering STO varactors highly reliable and repeatable below 10~K. Johnson's relation describes the resulting highly nonlinear dependence of $\epsilon_r$ on DC electric fields to first order and neglecting anisotropies and can be used to express the voltage-tunability of $C_{var}$ as \cite{Johnson1962, Ang2004, Apostolidis2020}:
\begin{equation}
    C_{var}\left(V_{var}\right) = \frac{C_0}{\left(1+\frac{(V_{var}-V_{offset})^2}{V_0^2}\right)^{1/3}}
    \label{CV_LGD}
\end{equation}
Here, $C_0$ is the zero-field capacitance, $V_{var}$ the applied varactor voltage and $V_{offset}$ accounts for an experimentally observed deviation of the peak capacitance from $V_{var}=\SI{0}{\volt}$. The voltage $V_0$ parametrizes the strength of the tunability of $\epsilon_r$ and is geometry-dependent \cite{Davidovikj2017}.

\begin{figure}[hbtp]
    \includegraphics[width=0.47\textwidth]{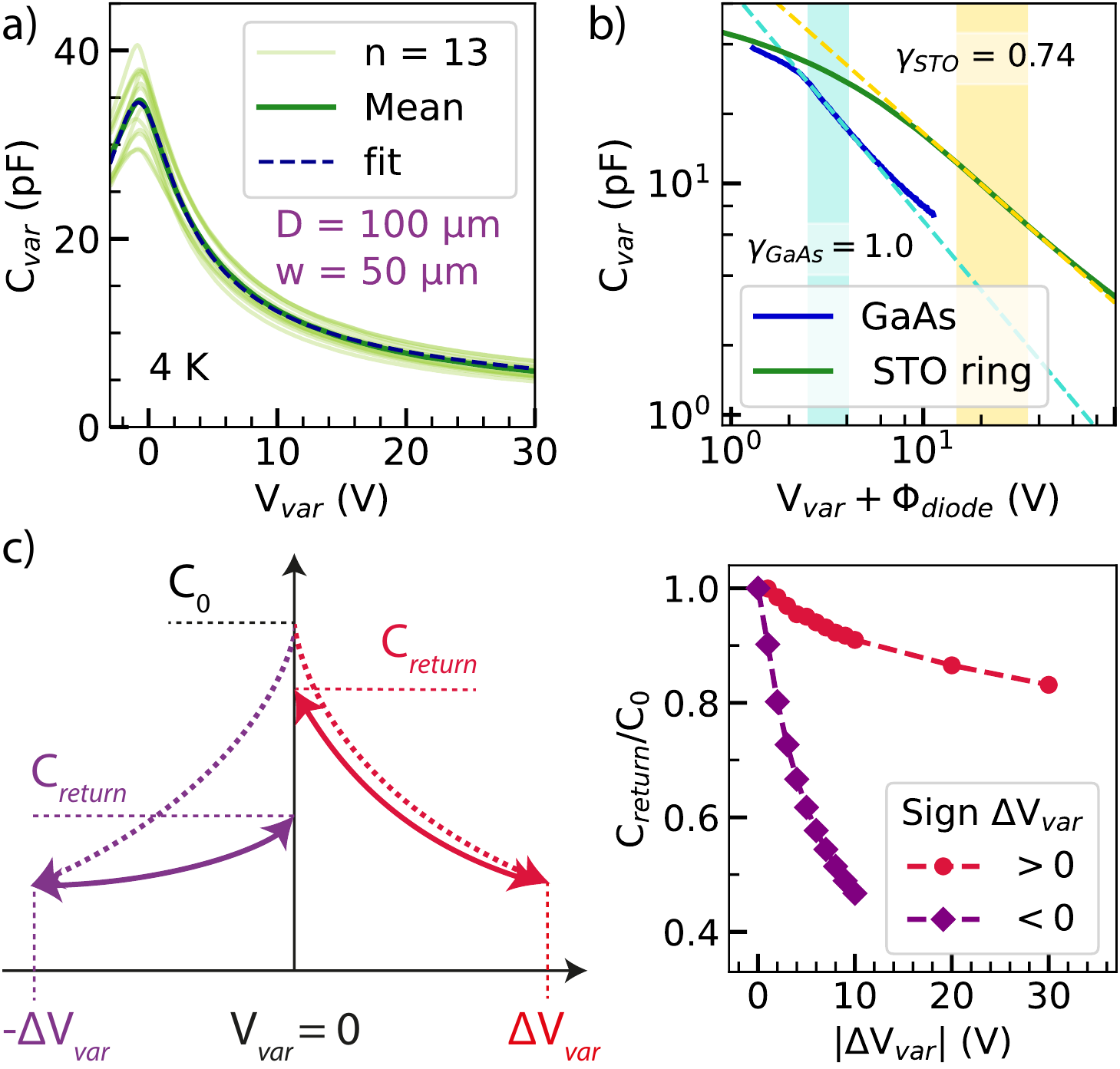}
    \caption{\textbf{CV-Characteristic} a) Varactor capacitance $C_{var}$ as a function of applied DC voltage measured on 13 nominally identical devices (light green), mean value (dark green) and fit of the mean to equation (\ref{CV_LGD}) (dark blue). Fitting all 13 traces yields $C_0=34\pm3\SI{}{\pico\farad}$, $V_{offset}=-0.8\pm0.1\SI{}{\volt}$ and $V_{0}=2.4\pm0.3\SI{}{\volt}$. b) Comparison of the capacitance-tunability of a commercial GaAs varactor with one exemplary STO ring at \SI{4}{\kelvin}. The yellow (turquoise) dashed lines are fits of equation (\ref{CV}) to the steepest regime of the STO (GaAs) CV-traces , indicated by the shaded areas. c) Schematic image (left panel) and quantitative analysis (right panel) of the varactor peak capacitance reduction. Applying negative $V_{var}$ (purple diamonds) causes a significant degradation of $C_{var}(V_{var}=\SI{0}{V})$ which is less severe for positive $V_{var}$ (red circles).}
    \label{fig:CV-Characteristic}
\end{figure}

From now on, we focus on devices with with $D=\SI{100}{\micro\meter}$ and $w=\SI{50}{\micro\meter}$ as these varactors covered the desired capacitance range and could be easily bonded. CV-characteristics of 13 nominally identical devices are shown in Fig. \ref{fig:CV-Characteristic} a) together with their average trace, indicating that all devices have capacitances within 10\% of the average. All curves show the characteristic maximum capacitance around zero $V_{var}$ and the suppression at larger voltages. Fitting all 13 traces to equation (\ref{CV_LGD}) reveals that the model agrees very well with the data over almost the entire range. Further, the three fit parameters are very similar for all devices, exhibiting a variability of only about 10\%.

For comparison with other varactors, we use the standard diode-varactor theory, which uses a power law exponent $\gamma$ as a benchmark for varactor performance, where a larger $\gamma$ indicates higher tunability. Abrupt varactors have $\gamma\leq0.5$ and if $\gamma>0.5$ the varactor is considered hyperabrupt, indicating highest tunability \cite{Buisman}. As a standard for comparison, a typical GaAs varactor used in similar experiments has $\gamma=1$ (Fig. \ref{fig:CV-Characteristic} b)) The CV-characteristic of a diode varactor near its peak tunability regime can be expressed as:
\begin{equation}
    C_{var}\left(V_{var}\right) = \frac{K}{\left(V_{var}+\Phi_{diode}\right)^\gamma}
    \label{CV}
\end{equation}

\noindent where $K$ is the capacitance constant, $V_{var}$ is the applied varactor voltage, $\Phi_{diode}$ is the built-in potential of the varactor \cite{Buisman}. For GaAs varactors, $\Phi_{GaAs}=\SI{1.3}{\volt}$ holds \cite{Heo2014}.

Fitting the peak tunability regime of the CV traces of all 13 devices to equation (\ref{CV}), we find hyperabrupt behavior over a large voltage range with mean $\overline{\gamma}_{STO}=0.72\pm0.05$ as shown in Appendix \ref{Supp_sect_gamma}. This is slightly higher than $\gamma\approx 0.66$ expected from (\ref{CV_LGD}). We note that in this high-tunability regime (yellow shaded in Fig. \ref{fig:CV-Characteristic} b)), the fit to equation (\ref{CV}) captures the measured capacitance slightly better than equation (\ref{CV_LGD}) as shown in Appendix \ref{Supp_sect_Power_v_TJhonson}. In analogy to the built-in potential, $V_{offset}$ is used as $\Phi_{STO}=\SI{0.8}{\volt}$ in the fits. To minimize the loss in tunability discussed in the following, we measure a single varactor from $V_{var}=\SI{0}{\volt}$ to $V_{var}=\SI{100}{\volt}$ as shown in \ref{fig:CV-Characteristic} b). The accessible capacitance range of the STO varactor is $C_{var}=\SIrange{45.0}{3.2}{\pico\farad}$, exceeding the range of a  commercial GaAs device used in comparable experiments (\textit{MACOM} MA46H204-1056) \cite{Ares2016,House2016,Ibberson2019} and of previous STO plate-capacitor implementations \cite{Apostolidis2020}. 

Interestingly, varactors with identical geometry but fabricated on $(100)$ and $(111)$ STO showed no significantly different CV-response as compared to the $(001)$ devices. While $(100)$ is equivalent to the $(001)$ direction, the zero-field dielectric constant of $(111)$ STO is $\epsilon_r=\num{12000}$ \cite{Neville1972}, half the value of $(001)$ STO. The approximately circular symmetry of the surface-electrode evenly integrates over all field directions in the crystal plane, effectively averaging away the anisotropy in $\epsilon_r$.

Sweeping from $V_{var}=\SI{0}{\volt}$ to a finite voltage $\Delta V_{var}$ and back to zero, we observe a reduction in $C_{var}(V_{var}=\SI{0}{\volt})$ from $C_{0}$ to $C_{return}$ (Fig. \ref{fig:CV-Characteristic} c)). If $V_{var}$ is subsequently kept within this same range, the CV-characteristic remains stable. The reduction is strongly dependent on the sign of $\Delta V_{var}$ as defined in Fig. \ref{fig:Geometry} a). Negative $\Delta V_{var}$ cause significantly bigger losses than positive $\Delta V_{var}$. This effect is quantified in the right panel of Fig. \ref{fig:CV-Characteristic} c), where $C_{return}/C_{0}$ is measured by sweeping to increasingly large $|\Delta V_{var}|$ and back to zero. The lost tunability range is fully recovered upon thermal cycling to $\SI{295}{\kelvin}$. Similar effects have been previously discussed with several paraelectric materials and can be explained by the polarization of microclusters in the crystal such as $O$-vacancies in STO \cite{Davidovikj2017,Ang2004}. This voltage polarity- and voltage history-dependent CV characteristic, while not related to ferroelectric hysteresis, has been referred to as hysteresis in the literature \cite{Davidovikj2017,Apostolidis2020}. The residual degradation of tunability present even when only applying positive voltages can easily be circumvented by choosing appropriate varactor dimensions and limiting the range of voltages applied.

\section{Gate-Dispersive Charge Sensing with Ge/Si core/shell Nanowire Double Quantum Dots}
 
\subsection{Impedance Matching}
\label{sect:matching}
 
In a next step, we implement gate-dispersive charge sensing with a Ge/Si core/shell NW hole DQD, using a STO varactor to optimize impedance matching. See Appendix \ref{Fab:NW} for NW device fabrication details. Quantum dots hosted in these NWs have been measured in direct current (DC) measurements \cite{Froning2018,Higginbotham2014a,Roddaro2008,Brauns2016,Zarassi2017,Hao2010}, using additional quantum dots as charge sensors \cite{Hu2007,Hu2011} and with an on-chip superconducting resonator \cite{Wang2019,Ungerer2022}. A strong, electrically tunable direct Rashba spin-orbit interaction (SOI) is present in Ge/Si core/shell NWs \cite{Kloeffel2011,Kloeffel2018,Froning2021,Hao2010}, enabling all-electrical spin control. Recent spin qubit experiments in this system have shown Rabi frequencies of several hundreds of MHz \cite{Froning2021a}.

\begin{figure}[hbtp]
    \includegraphics[width=0.47\textwidth]{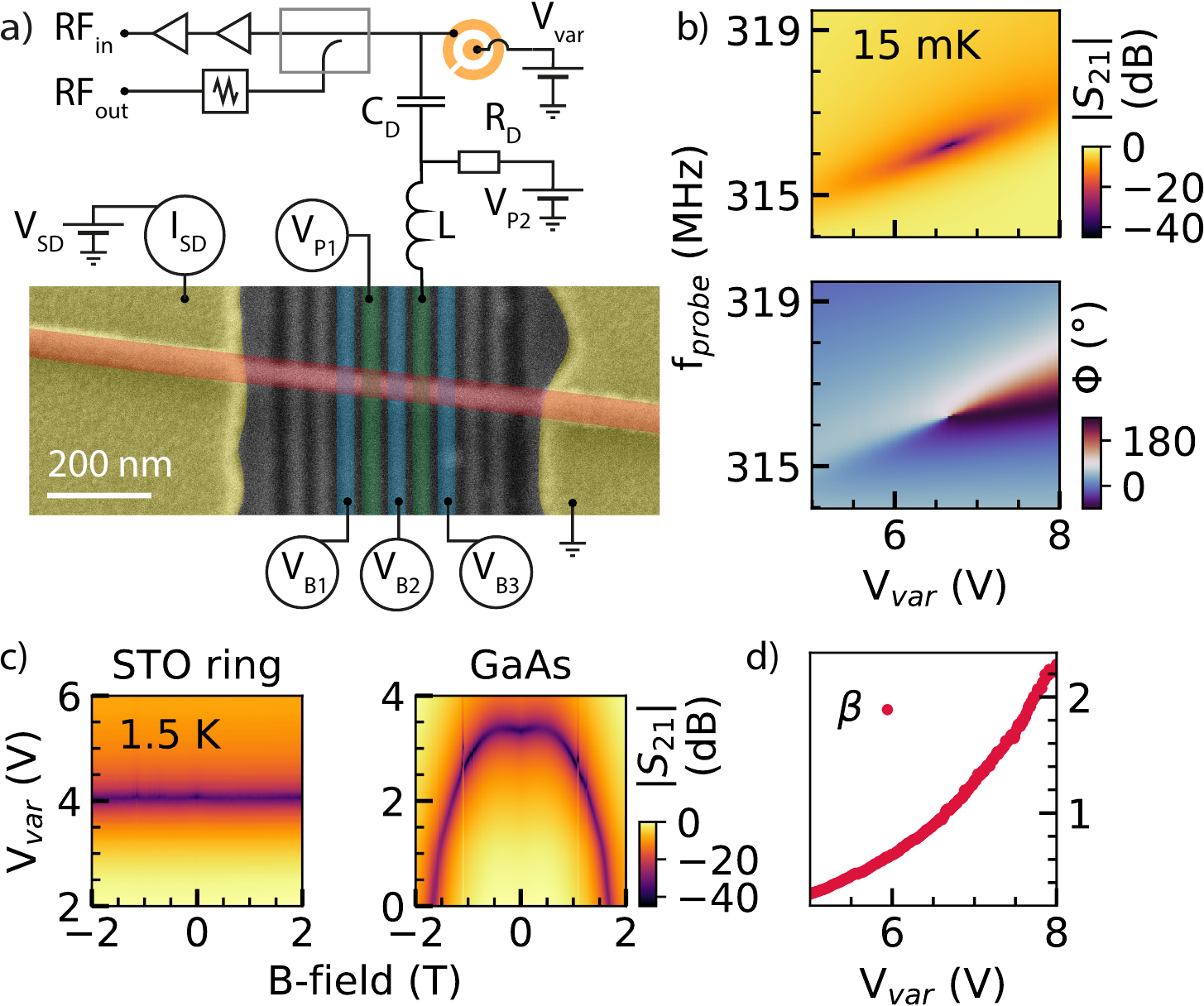}
    \caption{\textbf{Reflectometry setup and Impedance Matching} a) Gate-reflectometry circuit with STO matching-varactor. The false-colored SEM image shows a NW device with nine gates, five of which were used to form a DQD with two plungers (green) and three barrier gates (blue). The NW (red) is contacted on either side, enabling DC-current measurements through the wire. b) Impedance matching at probe temperature $T_{probe}\approx\SI{15}{\milli\kelvin}$. At perfect matching, the reflected amplitude is minimal and the phase response indicates a sudden shift from undercoupled ($V_{var}<\SI{6.625}{\volt}$) to overcoupled ($V_{var}>\SI{6.625}{\volt}$). c) Magnetic field-dependent impedance matching at $T\approx\SI{1.5}{\kelvin}$ without NW device. The tank with a STO ring varactor is unaffected by magnetic fields up to \SI{2}{\tesla}. Using a GaAs varactor, the matching voltage shifts strongly, complicating qubit measurements. d) Coupling coefficient $\beta$ as a function of $V_{var}$.}
    \label{fig:Setup}
\end{figure}

The reflectometry setup used to integrate an STO ring varactor with the NW device shown in Fig. \ref{fig:Setup} a) is described in more detail in Appendix \ref{Methods:Setup}. The tank resonance frequency at $f_{res}=316.20\pm0.05\,\SI{}{\mega\hertz}$ allows us to extract the parasitic capacitance $C_p=1.15\pm0.03\,\SI{}{\pico\farad}$ of the circuit. Tuning the tank as shown in Fig. \ref{fig:Setup} b), we reach impedance matching at probe temperature $T_{probe}\approx\SI{15}{\milli\kelvin}$ in a \textit{Bluefors} dilution refrigerator. We find the STO varactor to be highly temperature-resilient (Appendix \ref{Supp_sect_temp}). The resonator coupling coefficient $\beta=\frac{Q_0}{Q_{ext}}$ depicted in Fig. \ref{fig:Setup} d) is a convenient measure to evaluate the matching of the resonator impedance $Z_{res}$ to the external source circuitry impedance $Z_0$ and can be extracted from the $S_{21}$ response of the tank at resonance following Ibberson et al. \cite{Ibberson2019}. It is formally defined as the ratio of the unloaded ($Q_0$) and external ($Q_{ext}$) quality factors \cite{Ibberson2019}. The transition from the under- $(\beta<1;~Z_{res}>Z_0)$ to the overcoupled regime $(\beta>1;~Z_{res}<Z_0)$ is marked by a narrow minimum in the reflected amplitude $|S_{21}|$ and a sudden flip by $\SI{180}{\degree}$ in the phase of the reflected signal $\Phi$ where perfect matching $(\beta=1;~Z_{res}=Z_0)$ is reached.

The impact of an in-plane magnetic field on the matching with STO ring varactors and commercial GaAs varactors is compared in Fig. \ref{fig:Setup} c). Both scans were recorded at \SI{1.5}{\kelvin} because the GaAs varactor froze out at lower temperatures. No NW device was present for both scans. Fixing $f_{probe}$ near resonance at zero field, $B$ is subsequently swept against $V_{var}$. The GaAs varactor shows a strong dependence on $B$ even for $B\leq\SI{1}{\tesla}$, the typical spin qubit operation range \cite{Maurand2016,Hendrickx2021,Takeda2022,Camenzind2022,Froning2021}. The STO remains independent of B-field down to mK temperatures (Appendix \ref{Supp_sect_field}).

\subsection{Charge Sensing}
\label{sect:sensing}

\begin{figure*}[hbtp]
    \includegraphics[width=0.97\textwidth]{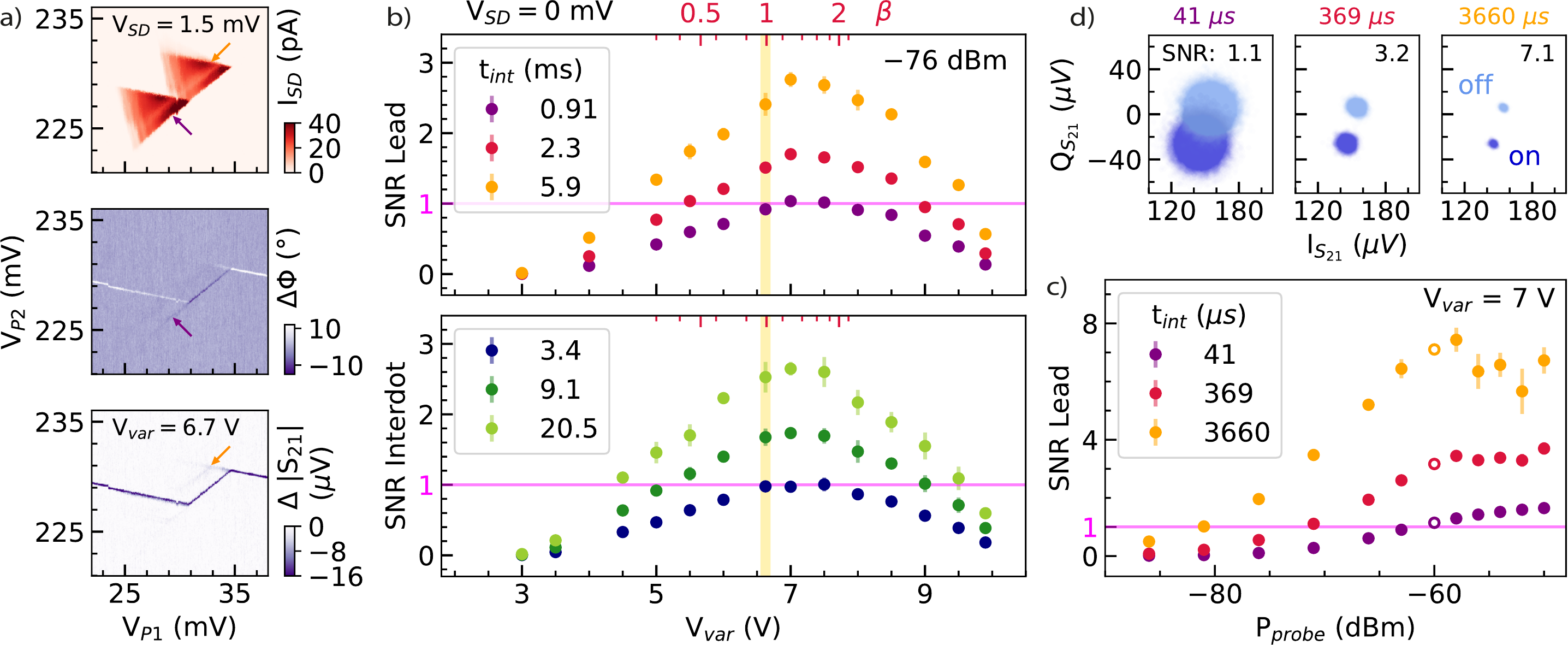}
    \caption{\textbf{Sensing a Ge/Si NW DQD} a) Charge stability map of a DQD formed in the Ge/Si core/shell NW, showing a characteristic bias triangle in the DC-current measurement (top). The simultaneously acquired reflectometry data (middle $\&$ bottom) feature an interdot line coinciding with the triangle baseline and lead transitions of the right dot with its reservoir. Orange arrows indicate the faintly visible excited state transition parallel to the baseline and purple arrows point to the baseline of the lower triangle. b) Time-domain SNR measurements of similar transitions as in a) as a function of $V_{var}$ and integration time $t_{int}$. Both transition SNRs behave similarly: The highest SNR regime is shifted to higher $V_{var}$ from the perfect matching (yellow shaded region). Investigating the interdot line, longer integration times are required to cross the SNR~$=1$ threshold (pink). Five independent SNR measurements were taken for each $V_{var}$. Error bars represent one standard deviation. c) The SNR can be increased by applying higher powers up to approximately $P_{probe}=\SI{-58}{\dBm}$ where the signal saturates. d) Demodulated data measured on (dark blue) and off (light blue) the lead transition at $P_{probe}=\SI{-60}{\dBm}$, corresponding to the empty circles in c).}
    \label{fig:Sensing}

\end{figure*}

Finally, we investigate charge transitions in a Ge/Si core/shell NW DQD using the impedance-matched tank circuit. Five bottom gates were used to define a DQD (Fig. \ref{fig:Setup} a)). Bias triangles, the hallmark signature of a DQD, are shown in Fig. \ref{fig:Sensing} a). Due to Coulomb blockade, source-drain current $I_{SD}$ can only flow through the DQD if at least one quantized energy level of each dot lies within the bias window. Estimating the plunger gate lever arms from the DC bias triangle yields $\alpha_{P1}\approx0.3$ and $\alpha_{P2}\approx0.24$ for the gates $P1$ and $P2$, respectively, in line with previous results on bottom-gated devices \cite{Froning2018}.

Gate-reflectometry allows for the measurement of tunneling between the dots (interdot transition) as well as between the dot under the sensor gate and a neighbouring reservoir (lead transition). Simultaneously measuring $I_{SD}$ and changes of the reflected RF amplitude $\Delta |S_{21}|$ and phase $\Delta \Phi$ demonstrates the correspondence of the two signals. In the reflectometry signal, the two reservoir transitions (white in $\Delta \Phi$) and the interdot transition (dark purple in $\Delta \Phi$) are clearly visible, appearing similarly in both $\Delta |S_{21}|$ and $\Delta \Phi$ signals. Note that the phase jumps by more than $|\Delta\Phi|=\SI{10}{\degree}$ at the transitions due to the close to perfect matching. In addition, the excited state appears faintly between the reservoir lines in parallel to the interdot transition (orange arrows), and a weak ground state transition of the lower triangle (purple arrows). The visibility of such features depends intricately on the balance of the tunnel barriers and the resulting tunnel rates in comparison to the resonator frequency. We note the remarkable stability of the quantum dots which can be formed in these NWs, manifesting in very sharp and highly repeatable transitions (see also Appendix \ref{Supp_sect_sensing}), only sporadically interrupted by a random charge switcher. A large-scale charge stability map is presented in Appendix \ref{Supp_sect_sensing} alongside investigations into the tunability of interdot tunneling using the barrier gate $B2$.

For the following analysis of lineshapes and SNRs, we measure at zero source-drain bias to reduce random charge switchers and maximize dispersive signal. We select a lead transition which shows up primarily in the reflected phase and a bright interdot transition near the triangle depicted in Fig. \ref{fig:Sensing} a). The lead transition lineshape can either be dominated by the reservoir hole temperature $T_H$, exhibiting a $\cosh^{-2}$-dependence on the gate voltage, or by tunnel-broadening which results in a Lorentzian lineshape with a tunnel rate $\gamma_L$ \cite{House2015}. Fitting the particular transition depicted in Appendix \ref{Supp_sect_lead}, we find slightly better agreement of the tail region with a $\cosh^{-2}$-relation, hinting towards the temperature broadened regime, i.e. that $k_BT_H>\hbar\gamma_L$ with $k_B$ Boltzmann's constant and $\hbar$ the reduced Planck's constant. We can thus extract $T_H\approx520\pm20\,\SI{}{mK}$. This temperature is far above the refrigerator temperature, presumably due to microwave irradiation from the cryogenic amplifier. The interdot tunnel coupling $t_c\approx22\pm5\,\SI{}{\micro\e\volt}$ can be extracted fitting the interdot transition shown in Appendix \ref{Supp_sect_interdot}. See Appendix \ref{Supp_sect_lineshapes} for estimates of the quantum capacitances associated with these transitions \cite{House2015}. Power broadening sets in above $P_{probe}\approx\SI{-60}{dBm}$ for both transitions mentioned, resulting in a reduced signal amplitude and larger width.

We investigate the impact of impedance matching through time-domain SNR measurements \cite{West2019} at the two transitions mentioned above. Note that a random charge switcher occured before we started the SNR measurements, resulting in a lower signal for the interdot transition than in the lineshape-investigation. We record time-traces of the demodulated reflectometry signal on and off a given charge transition for \SI{2}{\second} with a lock-in time constant $TC=\SI{4}{\micro\second}$. The in-phase ($I_{S_{21}}$) and quadrature ($Q_{S_{21}}$) components are then binned for a variable integration time $t_{int}$. Fig. \ref{fig:Sensing} d) depicts the two signals on the $IQ$-plane for different $t_{int}$. Comparing the distance of the two signals with the spread of the points estimates the SNR \cite{West2019}. 

Fig. \ref{fig:Sensing} b) shows the SNRs measured as a function of $V_{var}$. The need for longer integration at the interdot transition may be due to different tunnel rates. Both transitions exhibit the same qualitative dependence of SNR on $V_{var}$. The best SNR is in the slightly overcoupled regime at $V_{var}\approx\SI{7}{\volt}$ where $\beta=1.28$. In the case of dispersive sensing of a reactive load, the reduction of $C_{var}$ by increasing $V_{var}$ may lead to a higher tank quality factor, thus shifting the optimal sensing configuration away from perfect impedance matching \cite{Ibberson2019}. The observed shift of the tank resonance towards higher frequencies in \ref{fig:Setup} b) supports this hypothesis.  Evidently, the ability to tune impedance matching is of major importance for SNR optimization. Detuning the varactor by $\Delta V_{var}=\SI{2}{\volt}$ from the best operation point, causing a capacitance change of $\Delta\,C_{var}=\pm\,\SI{4}{\pico\farad}$, the SNR drops by up to a factor of $\sim 2$. 

The power-dependence of the SNR at the lead transition for $V_{var}=\SI{7}{\volt}$ is shown in \ref{fig:Sensing} c). We find that $P_{probe}>\SI{-60}{\dBm}$ causes significant instability of the system, as seen from the higher errors and plateauing of SNR, consistent with the onset of power broadening of the charge transitions (Appendix \ref{Supp_sect_lineshapes}). At $P_{probe}=\SI{-60}{\dBm}$, $t_{int}=\SI{41}{\micro\second}$ suffices to achieve SNR~$=~1.1$ (see Appendix \ref{Supp_sect_tc} for charge stability maps at low $TC$).

\section{Conclusion and Outlook}

In conclusion, we have fabricated and characterized hyperabrupt STO varactors with more than one order of magnitude tuning range. These varactors are not affected by moderate magnetic fields and can operate down to low mK temperatures.  The small size allows tight integration with other PCB components and scaling in applications with multiple tunable capacitors such as multiplexed readout. We demonstrate the high-frequency functionality of these varactors for gate-dispersive sensing of a Ge/Si core/shell NW hole DQD. Other charge-sensing schemes such as the RF-single-electron transistors or single-lead sensor dots should equally benefit from optimized impedance matching.

These results open the possibility for gate-dispersive single shot readout of spins in Ge/Si core/shell NWs and spin-relaxation measurements. Increasing the sensor gate lever arm as well as the resonator quality factor are important steps to improve the signal further. Expanding the varactor design, more elaborate matching networks including two tunable capacitors in one device may allow for greater flexibility and voltage-tunability of the tank resonance frequency. In particular, lowering $C_{max}$ while keeping $\gamma$ similar should enable impedance matching with high-Q superconducting inductors.

As a circuit element for the nascent field of cryogenic RF engineering, the hyperabrubt varactors can be used in a multitude of applications akin to their room-temperature diode counterparts. Such applications include voltage-controlled oscillators and frequency multipliers, essential components in scaled quantum computation architectures where cryogenic signal generation and manipulation will be necessary to overcome the wiring bottleneck \cite{Vandersypen2017}. More immediately, the STO varactors may be used in the ever-growing number of solid state systems where RF reflectometry techniques at cryogenic temperatures and in magnetic fields are employed such as hybrid devices featuring superconducting components, for fast thermometry or in materials science \cite{Vigneau2022}.

\begin{acknowledgements}
We thank A. Tarascio for his support during fabrication of the STO varactors. Furthermore, we acknowledge T. Berger for help with tank circuit modelling. We appreciated fruitful discussions with M. Hogg as well as his feedback on the manuscript. We thank G. Salis, E. Kelly and R. J. Warburton for useful discussions. Furthermore, we acknowledge S. Martin and M. Steinacher for technical support. This work was partially supported by the Swiss Nanoscience Institute (SNI), the NCCR SPIN, the Georg H. Endress Foundation, Swiss NSF (grant no. 179024), the EU H2020 European Microkelvin Platform EMP (grant no. 824109) and FET TOPSQUAD (grant no. 862046).
\end{acknowledgements}

\section*{Data availability}
The data supporting the plots of this paper are available at the Zenodo repository at \href{https://doi.org/10.5281/zenodo.7818326}{https://doi.org/10.5281/zenodo.7818326}.

\section*{Author Contributions}
R.S.E, T.P. and D.M.Z conceived of the project and planned the experiments. R.S.E. and D.T. fabricated and measured the STO varactors with inputs from T.P.. S.S. fabricated, tuned and operated the NW device. M.J.C. and P.C.K. helped with the NW device fabrication. R.S.E and S.S. integrated the varactor with the NW device and implemented the charge sensing experiments with inputs from T. P.. R.S.E, S.S., D.T. and T.P. analyzed the data with inputs from M.J.C., S.G., A.V.K. and D.M.Z.. A.L. grew the NWs under the supervision of E.P.A.M.B.. R.S.E. wrote the manuscript with inputs from all authors. D.M.Z. supervised the project.

\section*{Competing Interests}

The authors declare no competing interests.

\appendix

\section{Device Fabrication}
\subsection{Varactor fabrication}
\label{Fab:Var}

 The STO ring varactors were optically defined using a Heidelberg $\mu$PG laser-writer to pattern the structures on a commercially available $\rm{SrTiO}_3$ $(001)$ wafer (refractive index $n~=~2.39$) \cite{Apostolidis2020}. The STO chiplets used were \SI{0.5}{\milli\meter} thick, single-side polished, with a $\rm{TiO}_2$ termination on the back side, purchased from \textit{SurfaceNet}. A bilayer of photoresist, consisting of LOR 3A and S1805, was used. The laser-writer was operated in pneumatic focusing mode due to the optical transparency at the wavelength of \SI{405}{\nano\meter}. The structures were metallized with \SI{5}{\nano\meter}/\SI{65}{\nano\meter} of Ti/Au.  On one $10\times10~\SI{}{\milli\meter^2}$ chip, approximately $200$ devices were co-fabricated and cleaved into smaller segments with $5-15$ varactors for wirebonding and PCB integration. \\

\subsection{Nanowire device fabrication}
\label{Fab:NW}

 The QD device featured a set of $9$ bottom gates with a width of \SI{20}{\nano\meter} and a pitch of \SI{50}{\nano\meter}. The gates were fabricated by electron beam lithography (EBL) on an intrinsic Si $(100)$ chip with \SI{290}{\nano\meter} of thermal $\rm{SiO}_2$. Upon cold development \cite{Hu2004} the gates were metallized with \SI{1}{\nano\meter}/\SI{9}{\nano\meter} of Ti/Pd, respectively. In order to electrically insulate the gates from the NW, the gates were covered by a $\sim\SI{20}{\nano\meter}$ thick layer of $\rm{HfO}_2$ grown by atomic layer deposition. A single Ge/Si core/shell NW with core radius of $\sim\SI{10}{\nano\meter}$ and a shell thickness of $\sim\SI{2.5}{\nano\meter}$ \cite{ConesaBoj2017} was deterministically placed approximately perpendicular to the $9$ bottom gates. However, the exact in-plane angle is unknown. Finally, ohmic contacts were patterned by EBL and metallized with \SI{0.3}{\nano\meter}/\SI{50}{\nano\meter} of Ti/Pd after a \SI{10}{s} dip in buffered hydrofluoric acid to remove the native $\rm{SiO}_2$. The scanning electron micrograph presented in Fig. \ref{fig:Setup} shows a co-fabricated device.

\section{Reflectometry setup}
\label{Methods:Setup}

A STO varactor with $D=\SI{100}{\micro\meter}$ and $w=\SI{50}{\micro\meter}$ was integrated into a standard reflectometry setup shown in Fig. \ref{fig:Setup} a) on a printed circuit board (PCB). Attenuated coaxial cables were used to inject an RF tone at frequency $f_{probe}$ via a directional coupler (\textit{Mini-Circuits}  ZX30-17-5-S+) on the mixing chamber stage into a bias tee on the PCB ($C_D=\SI{87}{\pico\farad}$, $R_D=\SI{5}{\kilo\ohm}$). The varactor was operated as a tunable shunt-capacitor. A surface mount ceramic core inductor $L=220\pm6\,\SI{}{\nano\henry}$, in series with the parasitic capacitance of a bottom gate of the NW device formed the tank circuit. From the resonance frequency $f_{res}=316.20\pm0.05\,\SI{}{\mega\hertz}$ we find a parasitic capacitance $C_p=1.15\pm0.03\,\SI{}{\pico\farad}$. Filtered DC lines were used to provide gate-, source-drain bias- and varactor voltages. The reflected signal was amplified at the 4 K stage (\textit{Low Noise Factory} LNC0.2-3A) and at room temperature (\textit{B\&Z} BZY-00100700). A \textit{Zurich Instruments} UHFLI Lockin amplifier was used for signal generation and demodulation at room temperature. DC voltages were provided by a \textit{Basel Precision Instruments} SP927 digital to analog converter and $I_{SD}$ was amplified by a \textit{Basel Precision Instruments} SP983c current to voltage converter and recorded using an \textit{Agilent} 34410A digital multimeter.

\section{Varactor properties}

\subsection{Geometry}
\label{Supp_sect_geometry}

The results of COMSOL simulations of the $C_{var}$-dependence on $D$ and $w$ are shown in Table \ref{Tab1:C_Geometry}.

\begin{table}[htb!]

    \caption{\textbf{Simulated $C_{var}$ at $V_{var}=\SI{0}{\volt}$ in $pF$}}
    
    \begin{tabular}{c|r|r|r|r|r|r|r|r|}
    \multicolumn{1}{l|}{\textbf{\begin{tabular}[c]{@{}l@{}}$D \backslash w$\\ ($\mu$m)\end{tabular}}} & \multicolumn{1}{c|}{\textbf{10}}                  & \multicolumn{1}{c|}{\textbf{20}}                      & \multicolumn{1}{c|}{\textbf{30}}                   & \multicolumn{1}{c|}{\textbf{40}}                  & \multicolumn{1}{c|}{\textbf{50}}                     & \multicolumn{1}{c|}{\textbf{60}}                     & \multicolumn{1}{c|}{\textbf{70}} & \multicolumn{1}{c|}{\textbf{80}} \\ \hline
    \textbf{60}                                                                                       & 43                                                & 36                                                    & 34                                                 & 32                                                & 31                                                   & 31                                                   & 30                               & 30                               \\ \hline
    \textbf{80}                                                                                       & 62                                                & 52                                                    & 48                                                 & 45                                                & \begin{tabular}[c]{@{}r@{}}43\\ \end{tabular}    & \begin{tabular}[c]{@{}r@{}}42\\ \end{tabular}    & 42                               & 40                               \\ \hline
    \textbf{100}                                                                                      & 79                                                & \begin{tabular}[c]{@{}r@{}}68\\ \end{tabular}  & \begin{tabular}[c]{@{}r@{}}61\\ \end{tabular}  & 58                                                & \begin{tabular}[c]{@{}r@{}}56\\ \end{tabular}    & \begin{tabular}[c]{@{}r@{}}54\\ \end{tabular}    & 53                               & 52                               \\ \hline
    \textbf{120}                                                                                      & \begin{tabular}[c]{@{}r@{}}97\\\end{tabular} & 86                                                    & 75                                                 & 72                                                & 68                                                   & \begin{tabular}[c]{@{}r@{}}66\\ \end{tabular} & 65                               & 63                               \\ \hline
    \textbf{140}                                                                                      & 117                                               & 102                                                   & \begin{tabular}[c]{@{}r@{}}91\\ \end{tabular}  & 85                                                & \begin{tabular}[c]{@{}r@{}}81\\\end{tabular}    & \begin{tabular}[c]{@{}r@{}}79\\ \end{tabular}    & 77                               & 75                               \\ \hline
    \textbf{160}                                                                                      & 135                                               & \begin{tabular}[c]{@{}r@{}}120\\ \end{tabular} & \begin{tabular}[c]{@{}r@{}}106\\ \end{tabular} & \begin{tabular}[c]{@{}r@{}}99\\\end{tabular} & \begin{tabular}[c]{@{}r@{}}94\\ \end{tabular} & \begin{tabular}[c]{@{}r@{}}92\\ \end{tabular} & 89                               & 86                               \\ \hline
    \end{tabular}
\label{Tab1:C_Geometry}
\end{table}

Corresponding measurements at $T=\SI{4}{\kelvin}$ are shown in Table II. Note that only devices with $w\leq\SI{60}{\micro\meter}$ were fabricated and for small $w$ and $D$, measurements were often impossible because wirebonding created shorts between the two electrodes. Typical room-temperature capacitances were $\leq\SIrange[]{1}{1.5}{\pico\farad}$. In a few cases, the room-temperature capacitance was exceeding \SI{10}{\pico\farad}, likely originating from the DC wiring. These offsets were subtracted from the measured \SI{4}{\kelvin} data presented here. Note that two such cases were excluded from the variability analysis in the main paper. The devices used for the variability analysis were fabricated only after the results presented in Tables II and III were analysed. In order to remove the potential for systematic variations of the varactor geometry due to small variations in the fabrication parameters, the CV traces recorded for Tables II and III were excluded from the variability analysis.
The instances where two nominally identical devices show widely different capacitances at $V_{var}=\SI{0}{\volt}$ may be due to device damage from lift-off or wirebonding.

\begin{table}[hbtp]
\caption{\textbf{Measured $C_{var}$ at $V_{var}=\SI{0}{\volt}$ in $pF$}}
    \begin{tabular}{c|r|r|r|r|r|r|}
    \multicolumn{1}{l|}{\textbf{\begin{tabular}[c]{@{}l@{}}$D \backslash w$\\ ($\mu$m)\end{tabular}}} & \multicolumn{1}{c|}{\textbf{10}} & \multicolumn{1}{c|}{\textbf{20}} & \multicolumn{1}{c|}{\textbf{30}} & \multicolumn{1}{c|}{\textbf{40}} & \multicolumn{1}{c|}{\textbf{50}} & \multicolumn{1}{c|}{\textbf{60}} \\ \hline
    \textbf{80}                                                                                       & -                                & -                                & -                                & -                                & 36                               & 41                               \\ \hline
    \textbf{100}                                                                                      & -                                & 79,43                            & 52                               & -                                & 43                               & 46                               \\ \hline
    \textbf{120}                                                                                      & 45                               & -                                & -                                & -                                & -                                & 55,61                            \\ \hline
    \textbf{140}                                                                                      & -                                & -                                & 69                               & -                                & 78                               & 74                               \\ \hline
    \textbf{160}                                                                                      & -                                & 88,83                            & 80                               & 46                               & 74,27                            & 73,97                            \\ \hline
    \end{tabular}
\label{Tab2:C_0V}
\end{table}

To assess the tunability range, we present the measured values of $C_{var}$ at $V_{var}=\SI{30}{\volt}$ in Table \ref{Tab3:C_30V}.

\begin{table}[hbtp]
\caption{\textbf{Measured $C_{var}$ at $V_{var}=\SI{30}{\volt}$ in $pF$}}
    \begin{tabular}{c|r|r|r|r|r|r|}
    \multicolumn{1}{l|}{\textbf{\begin{tabular}[c]{@{}l@{}}$D \backslash w$\\ ($\mu$m)\end{tabular}}} & \multicolumn{1}{c|}{\textbf{10}} & \multicolumn{1}{c|}{\textbf{20}} & \multicolumn{1}{c|}{\textbf{30}} & \multicolumn{1}{c|}{\textbf{40}} & \multicolumn{1}{c|}{\textbf{50}} & \multicolumn{1}{c|}{\textbf{60}} \\ \hline
    \textbf{80}                                                                                       & -                                & -                                & -                                & -                                & 13                               & 12                               \\ \hline
    \textbf{100}                                                                                      & -                                & 24,18                            & 19                               & -                                & 14                               & 15                               \\ \hline
    \textbf{120}                                                                                      & 20                               & -                                & -                                & -                                & -                                & 21,21                            \\ \hline
    \textbf{140}                                                                                      & -                                & -                                & 25                               & -                                & 23                               & 25                               \\ \hline
    \textbf{160}                                                                                      & -                                & 26,72                            & 28                               & 15                               & 18,13                            & 29,26                            \\ \hline
    \end{tabular}
\label{Tab3:C_30V}
\end{table}

The limitations with regards to wirebonding should be easily overcome by adapting the electrode design to include remote, designated bondpads. This would make the smaller capacitance values listed in Table \ref{Tab3:C_30V} accessible without severe reduction of tunability. Furthermore $C_{var}\sim\SI{1}{\pico\farad}$ at $V_{var}=\SI{0}{\volt}$ may be reached if the dimensions of the ring are further reduced which would allow for different tank topologies and \cite{Ahmed2018} and tuning of the impedance matching with superconducting inductors.

\subsection{Fitting of the high-tunability range}
\label{Supp_sect_Power_v_TJhonson}
\begin{figure}[hbtp]
    \centering
    \includegraphics[width=0.47\textwidth]{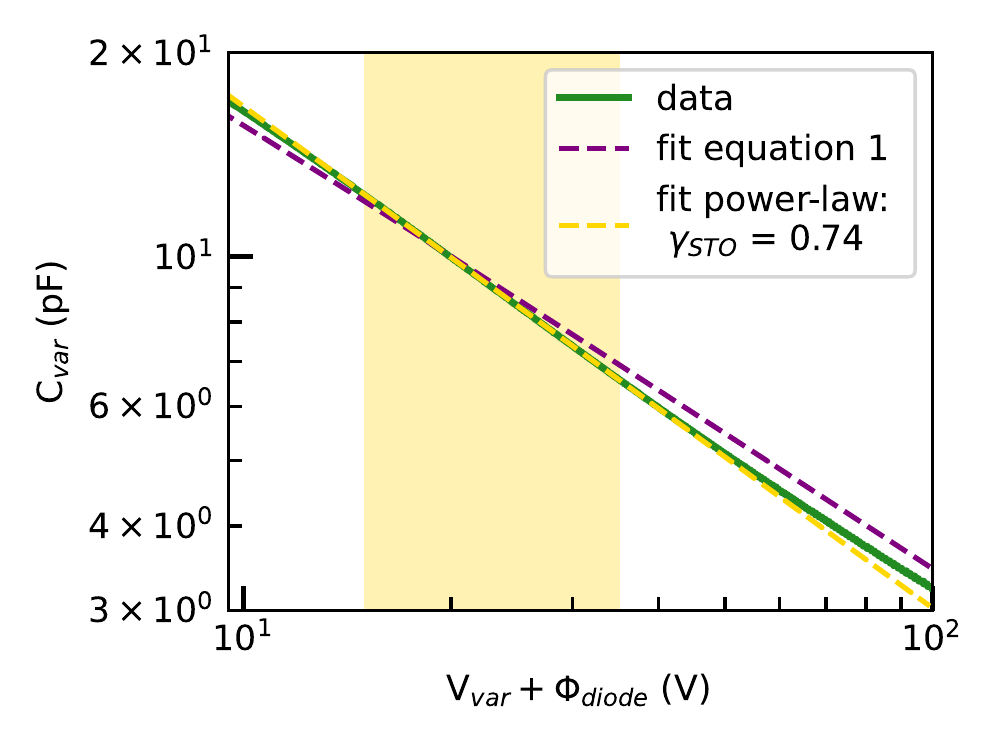}
    \caption{\textbf{Fitting the high-tunability regime:} Both fit functions show good agreement with the experimental data but the power-law from equation (\ref{CV_LGD}) with $\gamma = 0.74$ captures the relevant slope slightly better than Johnson's relation (\ref{CV}).}
    \label{Supp_fitting_jhonson}
\end{figure}

The two functions (\ref{CV_LGD}) and (\ref{CV}) can be used to fit the experimental CV trace. As shown in Fig. \ref{Supp_fitting_jhonson}, the agreement in the high-tunability regime is better for the power-law (\ref{CV}).

\subsection{Variability of hyperabrupticity}

\begin{figure}[hbtp]
    \centering
    \includegraphics[width=0.47\textwidth]{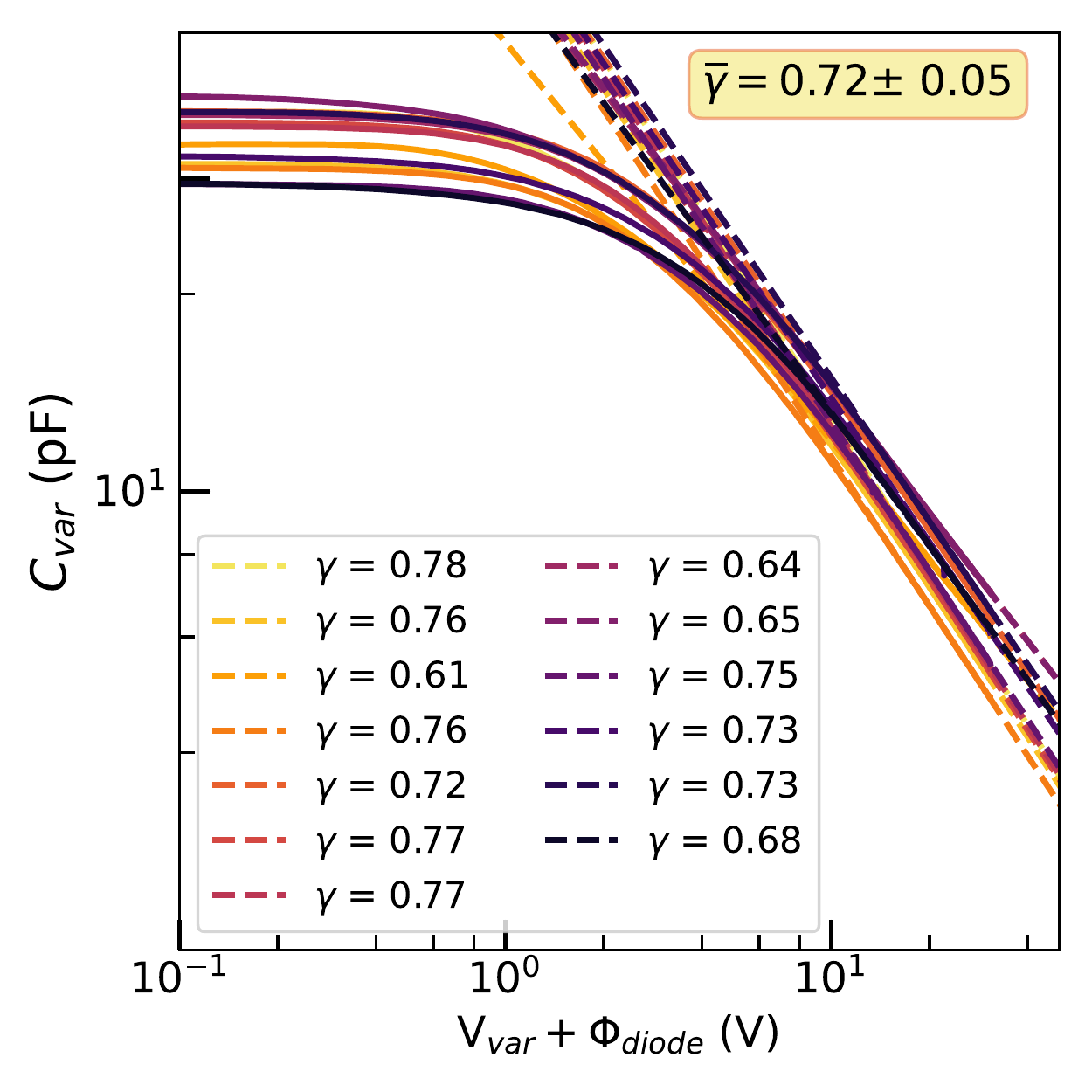}
    \caption{\textbf{Fits of CV-traces for all 13 devices from main Fig. 2 a)}} 
    \label{Supp_Gammas}
\end{figure}

All 13 devices depicted in main text Fig. \ref{fig:CV-Characteristic} a) show hyperabrupt CV-characteristics, as demonstrated in the fits shown in  Fig. \ref{Supp_Gammas}. We find the mean power-law exponent to be $\overline{\gamma}~=~0.72 \pm 0.05$.

\label{Supp_sect_gamma}

 \newpage
\subsection{Temperature-dependence of matching}

\label{Supp_sect_temp}
\begin{figure}[hbtp]
    \centering
    \includegraphics[width=0.47\textwidth]{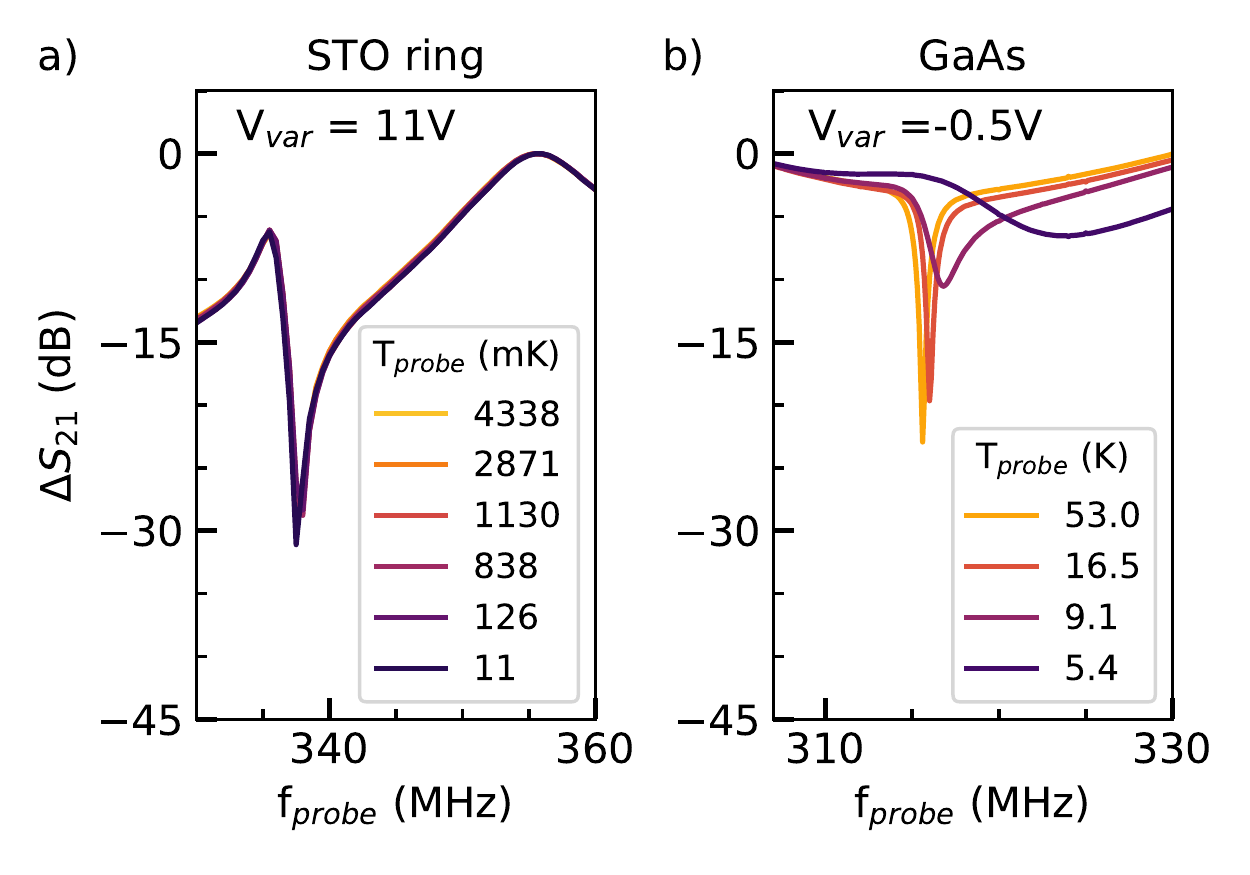}
    \caption{\textbf{Temperature-dependence of the tank resonance:} Two subsequent cooldowns of the sample PCB without NW device in the \textit{Bluefors} setup with a) an STO ring varactor and b) a commercial GaAs varactor used to tune impedance matching. } 
    \label{Supp_Temp}
\end{figure}

 Fig. \ref{Supp_Temp} shows the tank resonance recorded in the \textit{Bluefors} setup for a fixed $V_{var}$ using an STO varactor and, for comparison, using a GaAs varactor diode (\textit{MACOM} MA46H204-1056)\cite{Ares2016,Ahmed2018}. The STO varactor has almost no temperature-dependence below \SI{4.3}{\kelvin} down to \SI{11}{\milli\kelvin}. We have verified this robustness to temperature in multiple subsequent thermal cycles. No difference in varactor behavior was observed when tuning the varactor at \SI{}{\kelvin} temperatures and subsequently cooling the refrigerator to \SI{}{\milli\kelvin} versus keeping the varactor grounded while cooling to base temperature and tuning it later. The GaAs varactor in constrast shows a strong dependence on the temperature which sets in around \SI{53}{\kelvin}, making perfect matching impossible in the \SI{}{\milli\kelvin} regime.

\subsection{Zero-field anomaly}
\label{Supp_sect_field}
\begin{figure}[htb]
    \centering
    \includegraphics[width=0.47\textwidth]{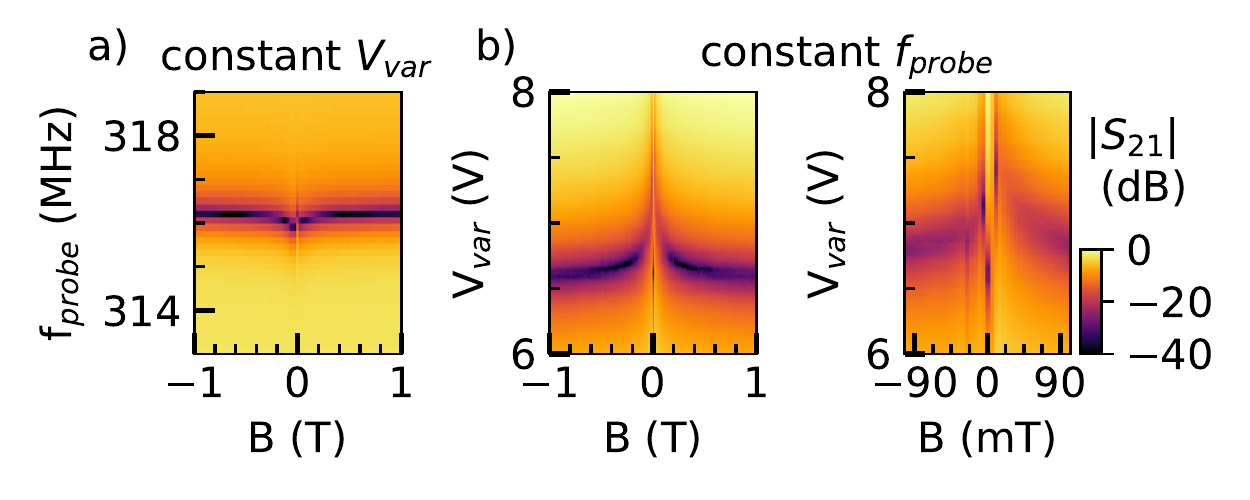}
    \caption{\textbf{Zero-field anomaly:} a) Shift of the tank resonance near zero field while $V_{var}$ is kept constant. b) The effect is visible much more prominent in a scan at constant $f_{probe}$, sweeping $V_{var}$. The right panel is a zoom-in on the low-field regime. Note the apparent discontinuities at $B<$\SI{10}{\milli\tesla}.} 
    \label{Supp_B0anomaly}
\end{figure}

In the \textit{Bluefors} LD setup, we observe a shift of the optimal matching condition within a range of $B\approx\pm\SI{50}{\milli\tesla}$ from $B=\SI{0}{\tesla}$ as shown in Fig. \ref{Supp_B0anomaly}. 

The origin of this shift remains elusive. Scans where $V_{var}$ is ramped rather than $f_{probe}$ show a stronger effect. The measurements presented in Fig. \ref{fig:Setup} show only a minor feature at $B=\SI{0}{\tesla}$. There are three apparent reasons for this discrepancy: 

First, it is possible that the STO varactor tunability/dielectric constant is affected in a different manner at $T\approx\SI{15}{\milli\kelvin}$ as compared to $T\approx\SI{1.5}{\kelvin}$. Since no such dependence was mentioned in previous work \cite{Apostolidis2020}, this appears to be a less likely explanation.

Second, the cryogenic amplifiers used in the two setups were different, namely a \textit{Cosmic Microwave} CLTF2 was used in the measurements at \SI{1.5}{\kelvin} (similar as CLTF1 used in \cite{Apostolidis2020}), but the setup at \SI{15}{\milli\kelvin} featured a \textit{Low Noise Factory} LNC0.2-3A amplifier. 

Third, the wirebonds used to bond the varactor are aluminium, whose normal-superconducting transition might show up around these fields. Further experiments are needed to finally exclude or confirm either hypothesis.

\section{Charge sensing}
\label{Supp_sect_sensing}

\subsection{Charge-stability maps}
\label{Supp_sect_overview}

\begin{figure}[hbtp]
    \centering
    \includegraphics[width=0.47\textwidth]{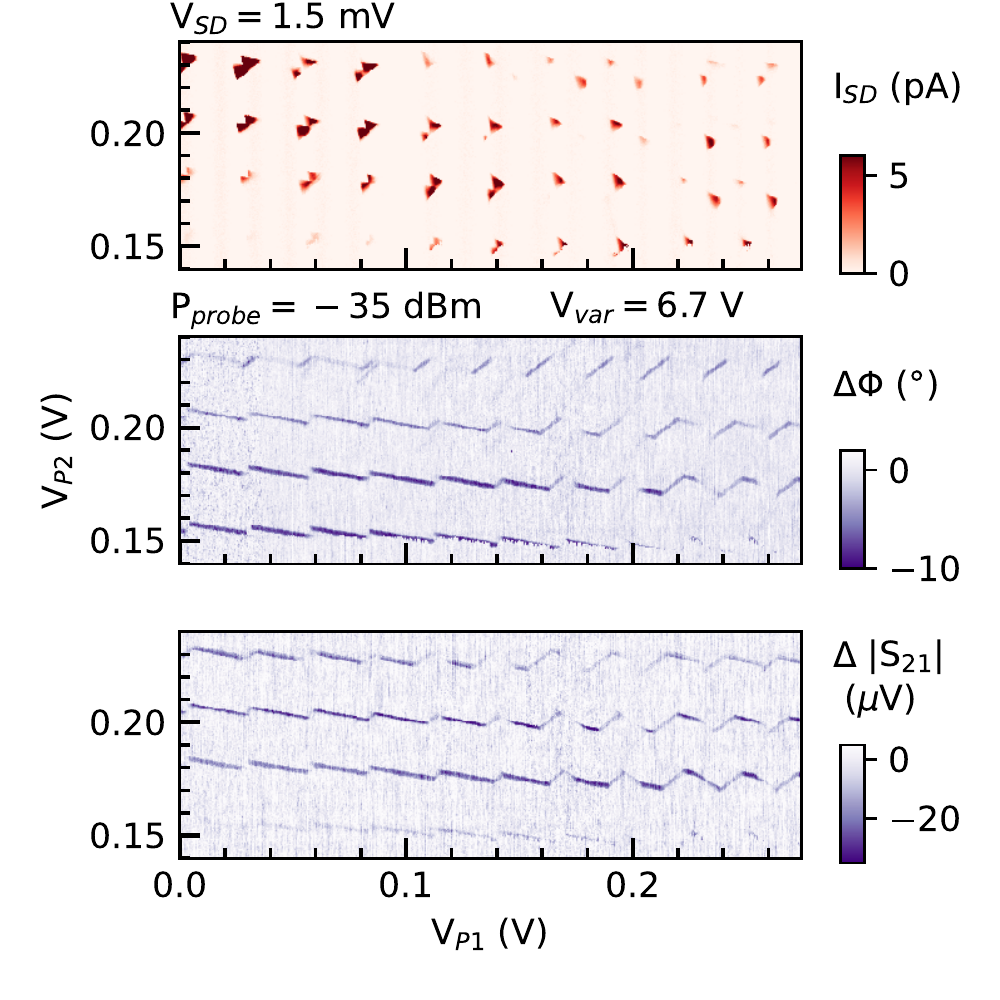}
    \caption{\textbf{Dispersive charge sensing over an extended plunger gate voltage range:} a) DC-current measurement of the charge stability diagram in the investigated Ge/Si core/shell nanowire device. b) Reflected phase $\Delta\Phi$ and c) RF amplitude $\Delta |S_{21}|$ measured along with the DC data in a). Here, we used a comparably large RF power $P_{probe}=\SI{-35}{\dBm}$, leading to significant power-broadening of the lead and interdot transitions and reducing the scan resolution requirements.}
    \label{Supp_overview}
\end{figure}

In Fig. \ref{Supp_overview} we present a larger scale charge-stability map of the Ge/Si core/shell nanowire DQD. We note that these data were taken at the same barrier gate voltages as the bias triangles presented in the main text. The DC-current signal shown in Fig. \ref{Supp_overview} a) exhibits multiple bias triangles, corresponding to transport in a double quantum dot (DQD). Towards the upper right quadrant of the graph, there is a region with signatures typical for a triple quantum dot (TQD) system, where the electrochemical potential of the center quantum dot is aligned with that of one of its adjacent outer quantum dots \cite{Schroeer2007}. The presence of both types of transport signatures in this configuration is an indicator for cotunneling involving the center quantum dot, as was previously observed in similar Ge/Si core/shell nanowire devices \cite{Froning2018}.

To make the lead and interdot transition lines more visible in the reflected signal, we increase the RF power to $P_{probe} = \SI{-35}{\dBm}$, which significantly broadens the transitions in the scan. In this specific charge stability diagram, we observe strong lead transitions at the regions where the DC transport signatures are those of a DQD, whereas the interdot transitions are more pronounced in the TQD region, indicating a change in the sensed quantum dot tunnel rates.

\clearpage

\subsection{Tuning the interdot tunnel rate}
\label{Supp_sect_interdot_tunnelrate}

\begin{figure}[hbtp]
    \centering
    \includegraphics[width=0.47\textwidth]{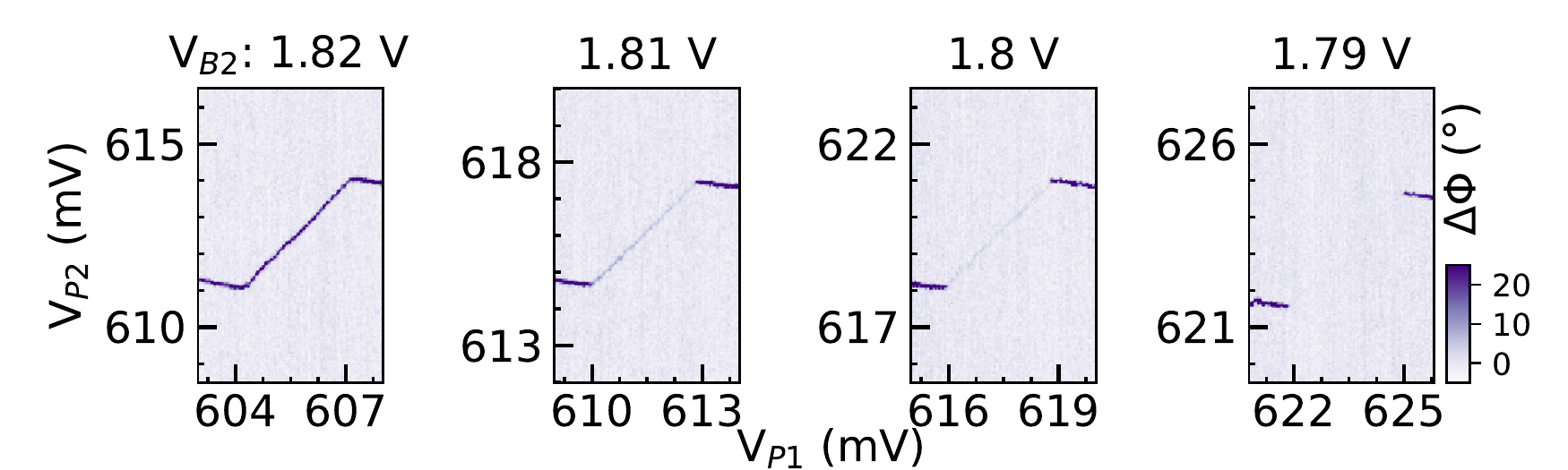}
    \caption{\textbf{Dependence on the center barrier gate voltage $V_{B2}$.} The strength of the reflected phase response $\Delta \Phi$ at the interdot line strongly changes within tens of mV applied to $V_{B2}$.} 
    \label{Supp_Interdot_Tuning}
\end{figure}

The strength of the charge sensing signal is dependent on the tunnel rates of the sensed charge transition. To illustrate this, we show series of charge stability maps in Fig. \ref{Supp_Interdot_Tuning}, tracking a particular interdot transition at $V_{SD}=\SI{0}{\milli\volt}$ for different center barrier gate voltages $V_{B2}$. As the voltage applied to the center gate is reduced, the measured interdot transition becomes more faint, until it fully vanishes at $V_{B2} = \SI{1.79}{\volt}$. The reduction in barrier voltage exponentially lowers the tunnel barrier, increasing the tunnel rate out of the sensitivity-window of our sensor \cite{Colless2013}.

\subsection{Reducing the lockin TC}
\label{Supp_sect_tc}

\begin{figure}[hbtp]
    \centering
    \includegraphics[width=0.47\textwidth]{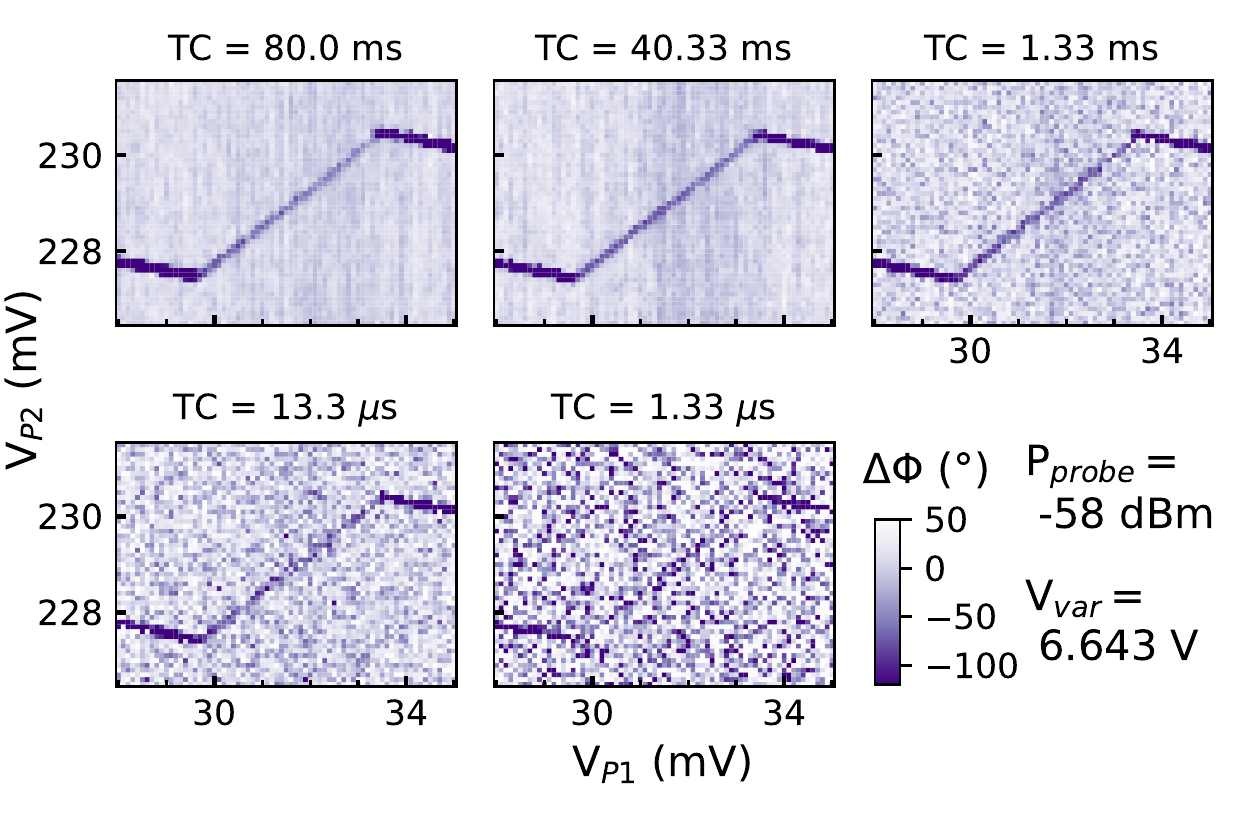}
    \caption{\textbf{Dependence on the integration time.} Interdot and lead transitions of a selected bias triangle, shown for varying $TC$. The signal can be recognized down to $TC = 1.33\mu$s.} 
    \label{Pushing_TC}
\end{figure}

In Fig. \ref{fig:Sensing} of the main text, we have determined the SNR for the lead and interdot transition of a selected bias triangle as a function of the integration time $t_{int}$. We plot the closeup of an interdot transition and adjacent lead transitions at varying $TC$ in Fig. \ref{Pushing_TC} to illustrate the possible measurement speedup. As $TC$ is decreased, the background noise increases, such that at $TC=\SI{1.33}{\micro\second}$ the transitions are barely visible anymore. 

\section{Charge-transition lineshape}
\label{Supp_sect_lineshapes}

Investigating the lineshape and linewidth of charge transitions, we can extract several quantum dot parameters. In the case of primarily dispersive signals, i.e. the transition manifests predominantly as a phase shift $\Delta \Phi$, we can model the signal as a capacitive correction to $C_p$, the quantum capacitance $C_q$. The dispersive shift $\Delta \Phi$ is translated to a change in circuit capacitance using the known slope $\frac{\delta \Phi}{\delta f}$ of $\Phi(f_{probe})$ at resonance for a given $V_{var}$ and the well-known frequency-relation for a resonant tank circuit \cite{House2015}:

\begin{equation}
    f_{res} = \frac{1}{2\pi\sqrt{LC}}
\end{equation}

\noindent where $f_{res}$ is the tank resonance frequency, $L$ the inductance and $C$ the capacitance of the resonator. The capacitive correction $C_q$ is then:

\begin{equation}
    C_q = \left(  2\pi\sqrt{L}\Delta\Phi\cdot\frac{\delta \Phi}{\delta f} +\frac{1}{\sqrt{C}}\right)^{-2}-C
\end{equation}

In the following, we present the results of lineshape investigations of the lead transition and the interdot transition used also for the SNR analysis in the main text.

\subsection{Lead transition}
\label{Supp_sect_lead}

\begin{figure}[hbtp]
    \centering
    \includegraphics[width=0.47\textwidth]{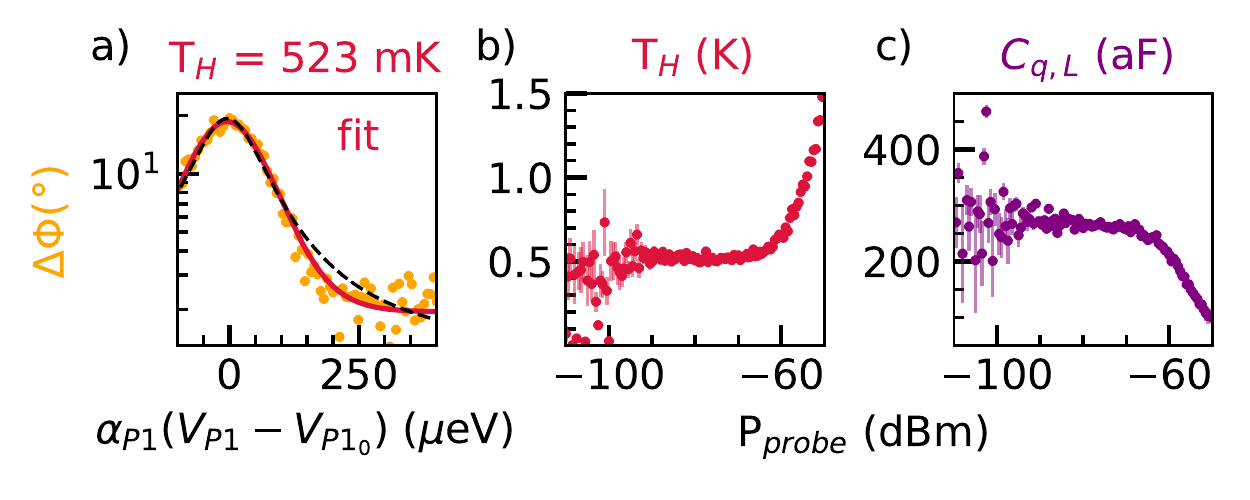}
    \caption{\textbf{Fitting the lead transition lineshape.} a) Example linecut (orange) of the lead transition recorded at $P_{probe}=\SI{-78}{\dBm}$. The $cosh^{-2}$ fit (red) captures well the flank of the peak where a Lorentzian model (dashed black) fails. Note the logarithmic y-axis scale and zoom-in on the right flank of the transition to highlight the discrepancy between the two models. b) Fitted hole temperatures $T_H$ as a function of probe power. c) Quantum capacitance of the lead transition $C_{q,L}$, calculated from the maximum phase shift at the charge transition. Power broadening causes a rise in the linewidth and reduced resonance height as observed for $P_{probe}>\SI{-60}{\dBm}$.} 
    \label{Supp_Lead_TH}
\end{figure}

Fitting the lead transition lineshape recorded at $V_{var}=\SI{7}{\volt}$, we find temperature-broadening to be dominant, $k_BT_H>\hbar\gamma_L$, where $k_B$ denotes Boltzmann's constant, $T_H$ the hole temperature of the lead, $\hbar$ the reduced Planck's constant and $\gamma_L$ the lead-dot tunnel rate \cite{House2015}. We fit the phase signal to:

\begin{equation}
    \Delta \Phi  = \Delta \Phi_{max} \cdot cosh^{-2}\left( \frac{\alpha_{P1} (V_{P1}-V_{P1_0})}{2 k_B T_H} \right)
\end{equation}

Here, $\Delta \Phi_{max}$ is the amplitude of the phase response and $V_{P1_0}$ the center of the peak along the $V_{P1}$ axis \cite{House2015}. 

The dependence of $T_H$ and $\Delta \Phi_{max}$ on the applied RF power $P_{probe}$ shown in Fig. \ref{Supp_Lead_TH} indicates power-broadening to set in above $P_{probe}=\SI{-60}{\dBm}$. In the regime of flat $T_H$, we find $T_H\approx520\pm20\,\SI{}{mK}$ and $C_{q,L}=266\pm8\,\SI{}{\atto\farad}$.

\subsection{Interdot transition}
\label{Supp_sect_interdot}

The dispersive signal at the interdot transition can be evaluated according to:

\begin{equation}
    \Delta \Phi  = \Phi_{0} \cdot t_c^2 \left( \left(\frac{ \alpha_{P1}(V_{P1}-V_{P1_0})}{2\sqrt{2}} \right)^2 +t_c^2 \right)^{-3/2}
\end{equation}

Where the constant prefactor $\Phi_0$ relates the term for the interdot quantum capacitance $C_{q,I}$ with the dispersive shift and $t_c$ is the interdot tunnel coupling \cite{House2015}. Note the additional factor $\sqrt{2}$ compared to \cite{House2015}, translating the quantum dot detuning axis into the plunger gate voltage axis.

\begin{figure}[hbtp]
    \centering
    \includegraphics[width=0.47\textwidth]{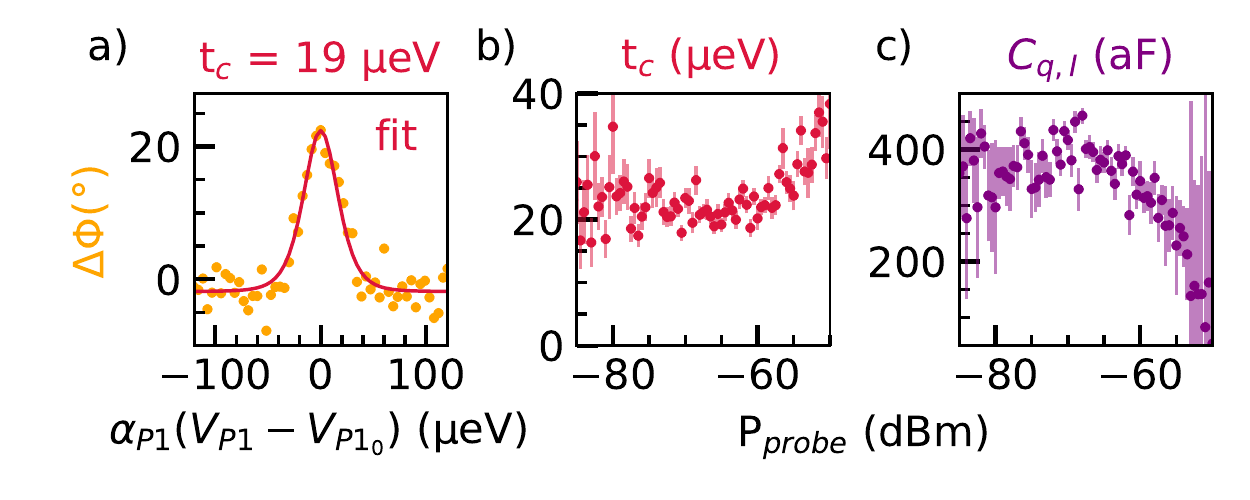}
    \caption{\textbf{Fitting the interdot transition lineshape.} a) Example linecut (orange) and fit (red) of the lead transition recorded at $P_{probe}=\SI{-66}{\dBm}$. b) Fitted tunnel coupling $t_c$ as a function of probe power. c) Quantum capacitance of the interdot transition $C_{q,I}$, calculated from the maximum phase shift at the charge transition. The signal is generally more unstable as compared to the lead transition, causing higher uncertainties.} 
    \label{Supp_Interdot_Tc}
\end{figure}

The results of fitting an interdot transition-linecut as a function of $P_{probe}$ are depicted in Fig. \ref{Supp_Interdot_Tc}. In the regime around $P_{probe}=\SI{-70}{\dBm}$, $t_c\approx22\pm5\,\SI{}{\micro\e\volt}$ and $C_{q,I}\approx380\pm70\,\SI{}{\atto\farad}$ are plateauing. Lower powers result in too low signal for reliable fitting while higher powers cause power broadening similar as in the case of the lead transition.
\clearpage
%

\end{document}